\documentclass{icrc2009}

\usepackage{graphicx}   
\usepackage{caption}  
\usepackage{fixltx2e}
\usepackage{url}

\newcommand{\shorttitle}[1]%
{\markboth{Proceedings of the 31\MakeLowercase{$^{st}$} ICRC, {\L}\'{o}d\'{z} 2009}{#1} }
\newcommand{\etal}{\MakeLowercase{\textit{et al. }}} 

\begin{document}
\title{Direct Measurement of the Atmospheric Muon Energy Spectrum with IceCube}

\author{\IEEEauthorblockN{Patrick Berghaus\IEEEauthorrefmark{1} for the IceCube Collaboration\IEEEauthorrefmark{2}}\\
\IEEEauthorblockA{\IEEEauthorrefmark{1}University of Wisconsin, Madison, USA}
\IEEEauthorblockA{\IEEEauthorrefmark{2}see special section of these proceedings}}

\shorttitle{P. Berghaus \etal IceCube Muon Energy Spectrum}
\maketitle

\begin{abstract}
Data from the IceCube detector in its 22-string configuration (IC22) were used to directly measure the atmospheric energy spectrum near the horizon. After passage through more than 10 km of ice, muon bundles from air showers are reduced to single muons, whose energy can be estimated from the total number of photons registered in the detector. The energy distribution obtained in this way is sensitive to the cosmic ray composition around the knee and is complementary to measurements by air shower arrays. The method described extends the physics potential of neutrino telescopes and can easily be applied in similar detectors. 
Presented is the result from the analysis of one month of IC22 data. The entire event sample will be unblinded once systematic detector effects are fully understood.
  \end{abstract}

 
\section{Introduction}

While the primary goal of IceCube is the detection of astrophysical neutrinos, it also provides unique opportunities for cosmic-ray physics \cite{Berghaus:2009jb}. One of the most important is the direct measurement of the atmospheric muon energy spectrum.

\begin{figure}
\begin{center}
\includegraphics[width=3.5in]{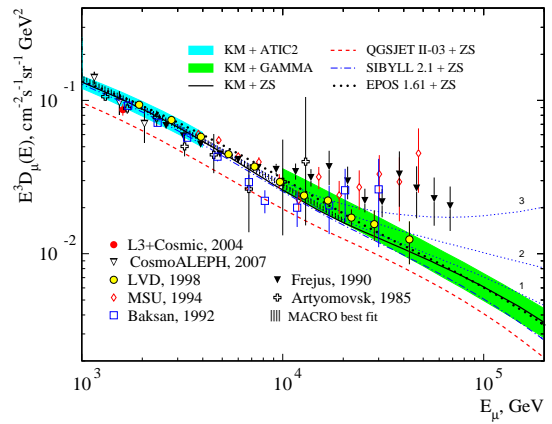}
\caption{Muon surface energy spectrum measurements compared to theoretical models \cite{Kochanov:2008pt}.}
\label{fig:yakutsk}
\end{center}
\end{figure}

As shown in figure \ref{fig:yakutsk} the energy spectrum of muons produced in cosmic-ray induced air showers has so far been measured only up to an energy of about 70 TeV \cite{Kochanov:2008pt}. The best agreement with theoretical models was found by the LVD detector, with the highest data point located at $E_{\mu}=\rm 40\ TeV$ \cite{Aglietta:1999ic}. All these measurements have been performed using underground detectors. Their sensitivity was limited by the relatively small effective volume compared to neutrino telescopes. 

With a planned instrumented volume of one cubic kilometer, IceCube will be able to register a substantial amount of events even at very high energies, where the flux becomes very low. The limitation in measuring the muon spectrum is given by its high granularity, and consequent inability to resolve individual muons. Most air showers containing high energy muons will consist of bundles with hundreds or even thousands of tracks. Since the energy loss per unit length can be described by the equation $dE/dx = a+bE$, low-energy muons will contribute disproportionately to the total calorimetric detector response, which depends strongly on the energy of the primary, disfavoring the measurement of individual muon energies.

This problem can be resolved by taking advantage of the fact that low energy muons are attenuated by energy losses during passage through the ice. In this analysis, the emphasis was therefore set on horizontal events, where only the most energetic muons are still able to penetrate the surrounding material. The primary cosmic ray interaction in this region takes place at a higher altitude, and therefore in thinner air. The reinteraction probability for light mesons (pions and kaons) is smaller and the flux of muons originating in their decays is maximized. 

The main possibilities for physics investigations using the muon energy spectrum are:

  \begin{itemize}
\item Forward production of light mesons at high energies. While muon neutrinos at TeV energies mostly come from the process $K \to \nu_{\mu} + X$, for kinematical reasons muons originate predominantly in pion decays $\pi \to \nu_{\mu} + \mu$ \cite{Gaisser:2002mi}. An estimate of the pion production cross section from accelerator experiments gives an uncertainty of 
$$\delta (\sigma_{N\pi}) \simeq 15\%+12.2\% \cdot log_{10}(E_{\pi}/500\: \rm GeV)$$
 at $x_{lab}>0.1$ above 500 GeV \cite{Barr:2006it}. This value should also apply in good approximation to the conventional (non-prompt) muon flux.

\item Prompt flux from charm meson decay in air showers \cite{Gelmini:2002sw}. Because of their short decay length, the reinteraction probability for heavy quark hadrons is negligible. The resulting muon energy spectrum follows the primary energy spectrum with a power law index of $\gamma \approx -2.7$ and is almost constant over all zenith angles. Since the non-prompt muon flux from lighter mesons is higher near the horizon, this means that the relative contribution from charm is lowest, and very challenging to detect.

\item Variations of the muon energy spectrum due to changes of the CR composition around the knee. Since the ratio of median parent cosmic ray and muon energy is $\leq 10$ at energies \cite{gaissenthomas} above 1 TeV, a steepening of the energy-per-nucleon spectrum of cosmic rays at a few PeV will have a measurable effect on the atmospheric muon spectrum at energies of hundreds of TeV. Comparison of the measured muon spectrum to various phenomenological composition models was the main focus of this analysis.

\end{itemize}

An additional benefit in the case of neutrino detectors is that a direct measurement of the muon flux will have important implications for neutrino analyses. By reducing the systematic uncertainties on atmospheric lepton production beyond 100 TeV, the detection potential for diffuse astrophysical fluxes will be enhanced. Also, atmospheric muons serve as a ``test beam'' that allows calibration of the detector response to high-energy tracks.

\section{Cosmic Ray Composition Models}

Starting from the hypothesis that most cosmic rays originate from Fermi acceleration in supernova shock fronts within our galaxy, the change in the energy spectrum can be explained by leaking of high energy particles. Since the gyromagnetic radius 

$$R=\frac{p}{eZB}\simeq (10pc)\frac{E_{prim}[PeV]}{ZB[\mu G]}$$

depends on the charge $Z$ of the particle, for a given energy nuclei of heavier elements are less likely to escape the galactic magnetic field than lighter ones.

The general expression for the flux of primary nuclei of charge $Z$ and energy $E_{0}$ is

$$\frac{d\Phi_{Z}}{dE_{0}}=\Phi{_Z^0}\left [ 1+\left (\frac{E_{0}}{E_{trans}}\right )^{\epsilon_{c}} \right ]^{\frac{-\Delta \gamma}{\epsilon_{c}}}$$

where the transition energy $E_{trans}$ corresponds to $\hat{E}_{p} \cdot Z$, $\hat{E}_{p} \cdot A$ or simply $\hat{E}_{p}$ for rigidity-dependent, mass-dependent and constant composition models. The parameter $\epsilon_{c}$ determines the smoothness of the transition, and $\Delta \gamma$ the change in the power law index.

Three alternative composition models have been proposed, which all can be fit reasonabkly well to the total cosmic ray flux in the region of the knee \cite{Hoerandel:2002yg}. These are:
  \begin{itemize}

    \item Rigidity-Dependent $\Delta \gamma$: This is the default composition used in the IceCube downgoing muon simulation. It is also the one favored by current models of cosmic ray production and propagation in the galactic amgnetic field.

    \item Mass-Dependent $\Delta \gamma$: An alternative model that also leads to a composition change around the knee. The change in the power law index does not depend on the charge, but on the mass of the nucleus. The best fit proposed in the original paper leads to a smaller value for the transition energy and a steeper spectrum after the cutoff. 

    \item Constant Composition: Here, the composition of the primary cosmic ray flux does not change. The knee is explained by a common steepening in the energy spectrum for all primaries occurring at the same energy. 

  \end{itemize}

\begin{figure}
\begin{center}
\includegraphics[width=3in]{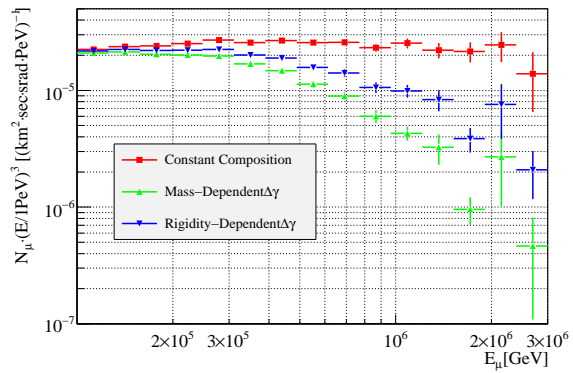}
\caption{Atmospheric muon energy spectrum at surface level averaged over the whole sky as simulated with CORSIKA/SIBYLL.}
\label{fig:musurf}
\end{center}
\end{figure}

The best measurement of the composition so far was done by KASCADE \cite{Antoni:2005wq}. Its result was consistent with a steepening of the spectrum of light elements, but depended strongly on the hadronic interaction model used to simulate the air showers (SIBYLL or QGSJET).

The influence of the three composition models on the muon energy spectrum is shown in figure \ref{fig:musurf}. While the spectrum for the constant composition model gradually changes from $E^{-3.7}$ to $E^{-4}$, the other two show a marked steepening corresponding to the cutoff in the energy per primary nucleon. By accurately measuring the muon energy spectrum, it is therefore possible to significantly constrain the range of allowed cosmic ray composition models in the knee region.

\section{Analysis}

The data set used in this analysis is based on the IceCube online muon filter, designed to contain all track-like events originating from the region below $70^{\circ}$. It covers the period from June 2006 to March 2007 with an integrated livetime of 275.6 days, during which IceCube was taking data with 22 strings (IC22). A number of quality cuts were applied in order to eliminate background from misreconstructed tracks and to reduce the median error in the zenith angle measurement to $\approx 0.7^{\circ}$. The final sample corresponded to an event rate of 0.146 Hz.

  \begin{table}[!h]
  \begin{center}
  \begin{tabular}{|c|c|c|c|}
  \hline
   $\theta_{zen} [deg]$  & $d_{vert}$ [km] & $d_{slant}$ [km] & $E_{\mu}^{thr} [TeV]$\\
   \hline
    0 & 1.5 & 1.5 & 0.28 \\
    70 & 1.5 & 4.39 & 1.12 \\
    70 & 2.5 & 7.31 & 2.59\\
    85 & 1.5 & 17.21 & 22.1\\
    85 & 2.5 & 28.68 & 207\\
  \hline
\end{tabular}	
  \caption{Threshold energy for muons passing through ice. The energy values correspond to an attenuation of 99.9\%.}
  \label{thresh_table}
  \end{center}
  \end{table}

To measure the single muon energy spectrum, it is necessary to reduce the background of high-multiplicity bundles, whose total energy depends primarily on the primary cosmic ray \cite{dima_2003}. Since there is no possibility to accurately estimate the multiplicity of a downgoing muon bundle, the only way to obtain single muons is by selecting a region close to the horizon to which muons of lower energies cannot penetrate. 

The minimum energy required for muons passing through a distance $d$ of ice can be approximated by the equation 

$$E_{cut}(d)=(e^{bd}-1)a/b$$

where $a=0.163\rm GeVm^{-1}$ and $b=0.192\cdot10^{-3}\rm m^{-1}$ \cite{Chirkin:2004hz}. The resulting threshold energies corresponding to vertical tracks and for tracks at the top and bottom of the detector for angles near the horizon are shown in Table \ref{thresh_table}.

Two factors determine the upper energy bound of this analysis. One is the contribution from atmospheric neutrinos, which will eventually dominate the event sample at large depths. The other, and more important, is the finite zenith angle resolution. Using simulated data, it was determined that it effectively limits the measurement of the slant depth to a values below 15 km. 

\begin{figure}
\begin{center}
\includegraphics[width=3in]{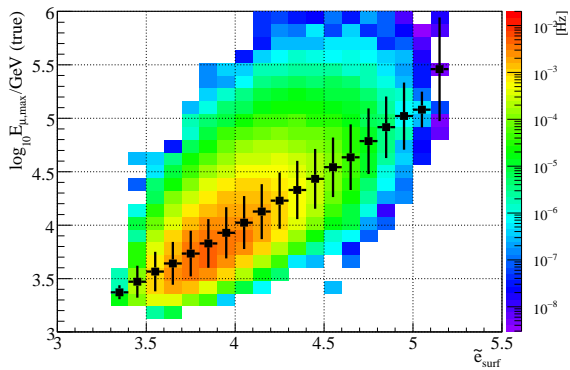}
\caption{Relation between energy proxy $\tilde{e}_{surf}$ and true surface energy of most energetic muon in shower. Here and in figure \ref{fig:mumult_byeps} the rigidity-dependent composition model was used.}
\label{fig:esurfprox}
\end{center}
\end{figure}

Using the slant depth alone, the range of this analysis is therefore insufficient to probe the region beyond 100 TeV. However, the reach can be extended by incorporating information about the energy of the muon as it passes through the instrumented volume. 

For muon tracks in the detector, the energy resolution approaches $\Delta \log_{10}(E) \approx 0.3 $ above 10 TeV \cite{juande_dima}. This information can be combined with the slant depth to obtain a better estimate for the muon energy at the surface.

A natural way to do this is by defining a surface energy proxy $\tilde{e}$ that behaves as

$\exp(\tilde{e}_{surf}) \propto \log{n_{\gamma}} \cdot d_{slant}$

where $n_{\gamma}$ represents the total number of photons measured by the detector. Figure \ref{fig:esurfprox} shows the resulting parameter, which has been linerly rescaled in such a way that its value corresponds to the mean $\log(E_{\mu,surf}/GeV)$ for any given bin, provided that the muon energy spectrum is reasonably close to the standard $E^{-3.7}$.

\begin{figure}
\begin{center}
\includegraphics[width=3in]{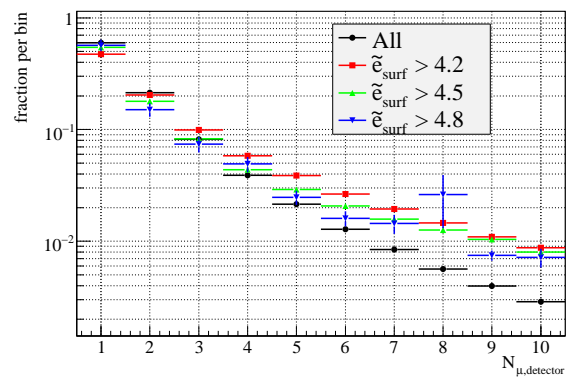}
\caption{Simulated muon multiplicity for atmospheric showers at closest approach to the center of the InIce detector for different values of $\tilde{\epsilon}$.}
\label{fig:mumult_byeps}
\end{center}
\end{figure}

\begin{figure}
\begin{center}
\includegraphics[width=3in]{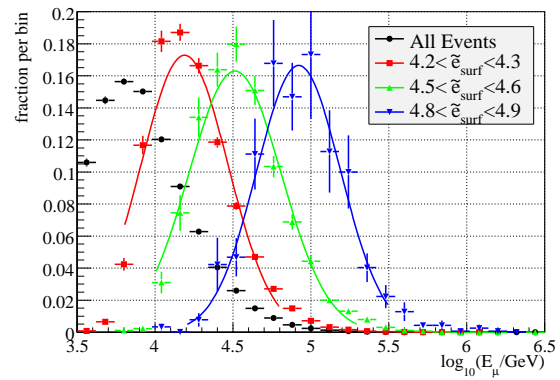}
\caption{True surface energy of most energetic muon in shower using rigidity-depending composition model for different values of $\tilde{e}_{surf}$, with fits to Gaussian function. All individual distributions are normalized to unity.}
\label{fig:rashianity}
\end{center}
\end{figure}

An important criterion for the applicability of the energy proxy parameter is that over the entire range of measurement the muon multiplicity remains low, and the influence of high-multiplicity bundles small. Figure \ref{fig:mumult_byeps} confirms that this is indeed the case. It should be noted here that the most energetic muon typically accounts for the dominant contribution to the total energy in the detector, such that other tracks in the bundle can be neglected.

The spread in muon surface energies for a given value of $\tilde{e}_{surf}$ is shown in Figure \ref{fig:rashianity}. Around the peak the distributions can be approximated by a Gaussian whose width lies in the range of $\Delta \log_{10}E \approx 0.3-0.4$.

\begin{figure}
\begin{center}
\includegraphics[width=3in]{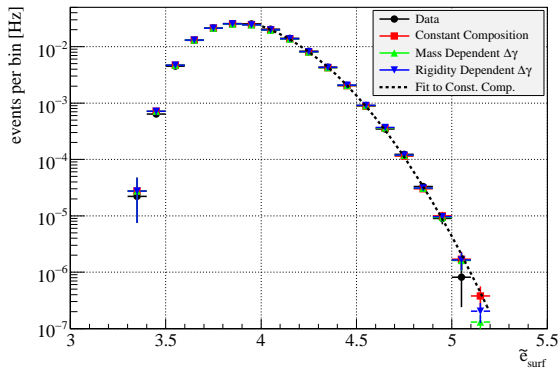}
\caption{Data from one month of IC22 at final cut level compared to simulated event rates and fit of empirical function $exp(a+b\tilde{e}+c\tilde{e}^{2})$ to constant composition distribution.}
\label{fig:3mod_fit}
\end{center}
\end{figure}

\begin{figure}
\begin{center}
\includegraphics[width=3in]{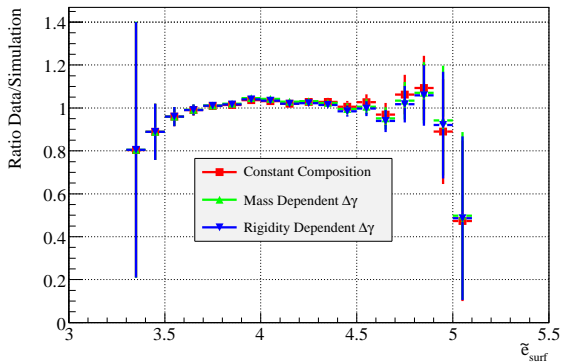}
\caption{Ratio of experimental to simulated $\tilde{e}_{surf}$ distributions for one month of IC22 data. Uncertainties are statistical only and exclude systematic detector effects.}
\label{fig:datsimrat}
\end{center}
\end{figure}

\section{Result}

Figure \ref{fig:3mod_fit} shows simulated event rates in dependence of $\tilde{e}_{surf}$ compared to data at final cut level. Almost over the entire range all three models can be approximated by the same empirical fit function. Only in the highest bin can a distinction be made.

The experimental data agrees remarkably  well with the simulation, as can be seen more clearly in Figure \ref{fig:datsimrat}. Despite the steeply falling distribution, the ratio of data to simulation remains very close to one over almost the entire range. For $\tilde{e}>5$, corresponding to $E_{\mu}>\rm 100\ TeV$, the measurent is based on only three data events.

Using the entire year of IC22 data, the predicted event yield for $5.1<\tilde{e}_{surf}<5.2$ based on the constant composition model corresponds to about 10 events. It is therefore unlikely, even neglecting systematic uncertainties, that any of the three models under consideration could definitely be excluded yet. This situation is expected to change as soon as 40-string data can be included in the analysis.

\section{Conclusion}

This result demonstrates the potential for an accurate measurement of the muon energy spectrum with large neutrino detectors. So far only one month of data has been considered in the analysis, corresponding to about 10\% of the entire event sample. Nevertheless, the measurement already covers an energy range almost a factor of three above that of the previous upper limit, with very good agreement between data and simulation.

While it will be difficult to make a definitive statement about the cosmic ray composition around the knee based on IC22 data, it will be possible to confirm the validity of cosmic ray air shower models up to previously inaccessible energy ranges. 

At the time of writing, the instrumented volume of the detector has increased by a almost factor of three. Further enlargements are scheduled for the next few years. Future measurements of the muon energy spectrum will benefit from a larger effective area, and a substantial improvement in the angular resolution related to the longer lever arm for horizontal muon tracks within the detector.

Once residual systematic detector uncertainties are resolved, a comprehensive analysis that accounts for both the energy spectrum of individual muons and the total shower energy in the detector will be feasible. The potential for such a combined measurement is unique to large volume detectors.


\end{document}